\documentclass[multphys,vecphys]{svmult}

\usepackage{makeidx}         
\usepackage{graphicx}        
\usepackage{multicol}        
\usepackage[bottom]{footmisc}

\makeindex             
\begin{document}

\title*{The optical and near-infrared properties of nearby groups of galaxies}
\titlerunning{Optical and near-IR properties of nearby groups}
\author{Somak Raychaudhury \and Trevor A. Miles}
\authorrunning{Raychaudhury \& Miles}
\institute{School of Physics \& Astronomy, University of Birmingham, UK}
\maketitle
%
\begin{abstract}
We present a study of the optical (BRI) and near-infrared (JHK)
luminosity fuctions (LFs) of the GEMS sample of 60 nearby groups of
galaxies between $0.01\!<\! z \!<\! 0.04$, with our optical CCD photometry
and near-IR photometry from the 2MASS survey.  The LFs in all
filters show a depletion of galaxies of intermediate luminosity, two
magnitudes fainter than $L_\ast$, within $0.3\,R_{500}$ from the
centres of X-ray faint groups. This feature is not as
pronounced in X-ray bright gropus, and vanishes when LFs are found out
to $R_{500}$, even in the X-ray dim groups.  We argue that this
feature arises due to the enhanced merging of intermediate-mass
galaxies in the dynamically sluggish environment of low
velocity-dispersion groups, indicating that
merging is important in galaxy evolution even at $z\!\sim\! 0$.
\end{abstract}

\section{Introduction}
\label{sec:intro}
Observational studies of the environmental dependence of galaxy
evolution mostly concentrate on galaxies in rich clusters, even
though $<5$\% of galaxies are found in such environments. Most
galaxies are found in groups, and arguably the group environment plays
an important role in their evolution. In groups, galaxies could be
transformed by their interaction with the intragroup medium, or with
other galaxies by means of a variety of processes (e.g. stripping,
tidal interaction) or through  merging with other galaxies. 
Here we seek to study the optical and near-IR properties of galaxies
in groups to find whether they support the relative importance of any
of these processes.

\begin{figure}
\centering
\includegraphics[height=5.3cm,angle=-90]{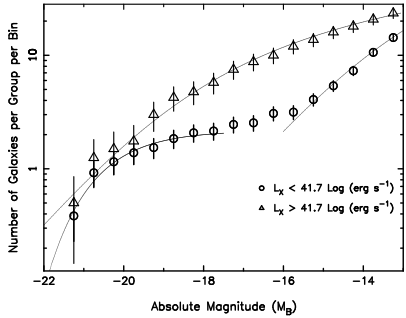}
\caption{Cumulative $B$-band Luminosity Function of 25 GEMS 
galaxy groups, divided into two categories: 
X-ray bright groups ($L_X>10^{41.7}$ erg~s$^{-1}$, plotted as 
triangles and  X-ray dim groups ($L_X<10^{41.7}$ erg~s$^{-1}$), plotted
as circles. Only the LF of X-ray bright groups can be
fit with a single Schechter function.
}
\label{fig:cumulf}       
\end{figure}

\section{Luminosity functions of GEMS groups}
\label{sec:2}

We explore the optical properties of galaxies in a sample of 60
nearby groups, known as the Group Evolution Multi-wavelength Study
(GEMS\index{GEMS}, 
detailed descriptions in \cite{op04,forbes06}.
This sample represents a variety of groups over a large range
of evolutionary stages, and are all in the fields of $>10$~ks of ROSAT
PSPC observations (some of them are not detected).  As a description
of the Group environment, we use their bolometric X-ray luminosity
$L_X$, and divide the sample into two subsamples: X-ray bright if
$L_X\!>\!10^{41.7}$ erg~s$^{-1}$, and X-ray dim if less
(including the undetected ones). This X-ray
luminosity refers to that of the group plus any central galaxy that
might exist (for more details, see \cite{miles04}.).

\subsection{B(VR)IJHK photometry}

The optical subsample consists of 25 GEMS groups: 17 of them
were observed at the 2.5m INT, La Palma, with the WFC, imaging
an area of $4\times 22.5\times 11.3$ arcmin of sky with BVI filters. 
Another 8
groups were observed with the 2.2m ESO/MPI telescope at La~Silla,
Chile, using the WFI, with a field of $34\times
33$ arcmin, with broadband BRI filters. For each group, we went out to
a radius of $0.3\, R_{500}$ from their centres.

Furthermore, we extracted JHK magnitudes of all 60 GEMS groups from
the 2MASS All-Sky Extended Source Catalog (2MASX)\index{2MASS}, 
going out to a
radius $R_{500}$ of its centre for each group, down to a limiting
magnitude of $M_K\!=\!13.75$.  The adopted group centres, values of
$R_{500}$ and distances to these groups can be found in \cite{op04}, 
and details of member selection and data reduction in \cite{miles04,miles06a}.

\begin{figure*}
\centering
{\includegraphics[height=5.4cm,angle=-90]{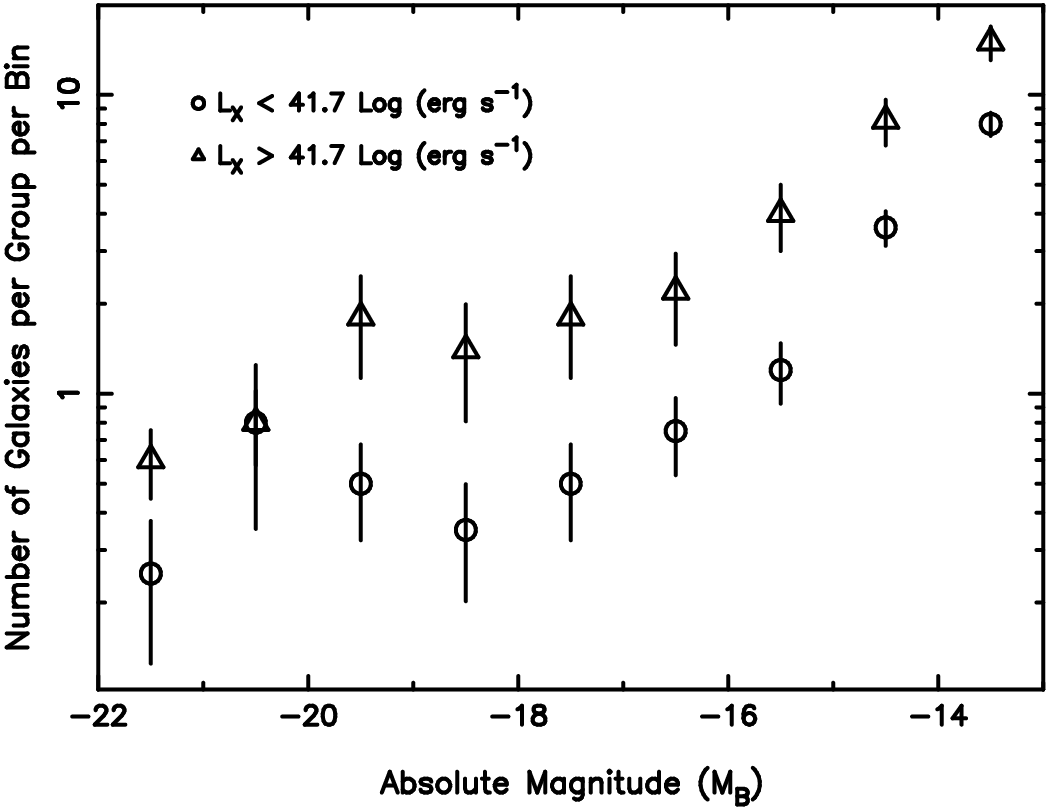}
\includegraphics[height=5.4cm,angle=-90]{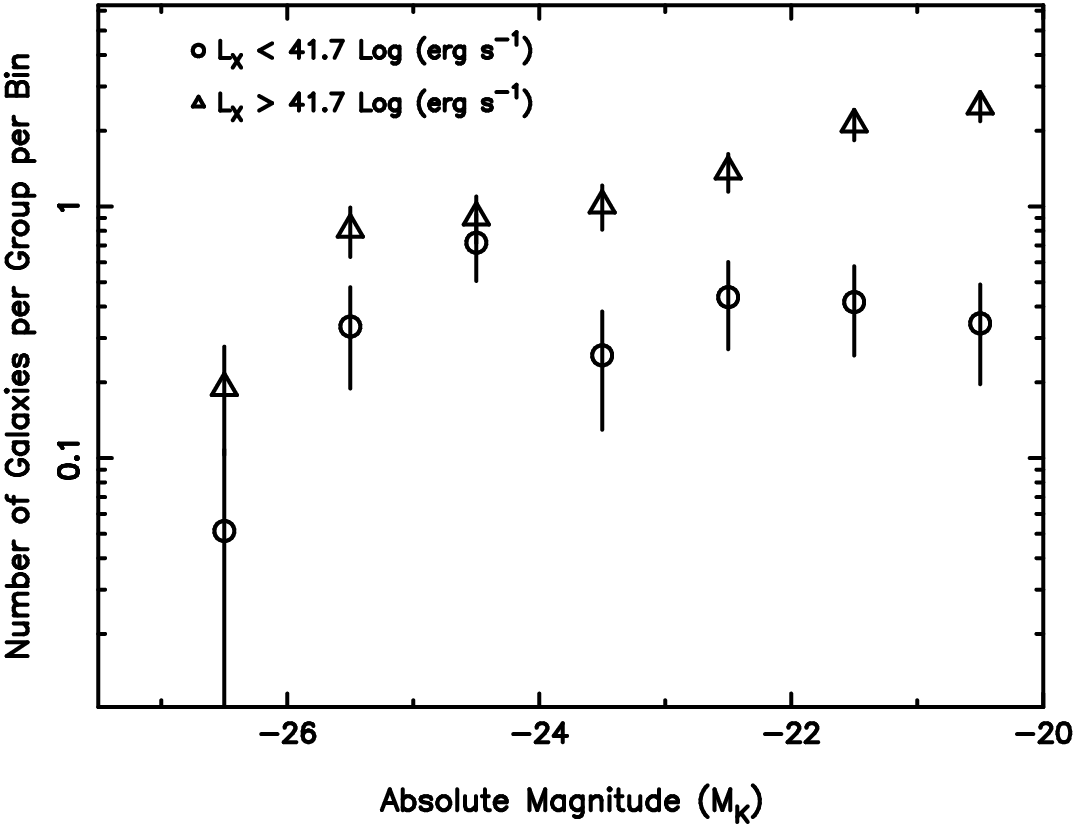}}
\caption{ The mean differential Luminosity functions of GEMS Groups,
within 0.3$R_{500}$ from their respective centres, {\it
(Left:)} in the $B$-band, for a subsample of 25 groups, and {\it (Right:)}
$K$-band (2MASS), for all 60 GEMS groups.  They are divided into X-ray
bright groups ($L_X\!>\!10^{41.7}$ erg~s$^{-1}$, triangles) and X-ray dim
groups (circles), stacked together to
form composite LFs for the respective sub-classes.  
The LFs of the dim groups show ``dips'' between
$-19\!<\!M_B\!<\!-17$ and $-24\!<\!M_K\!<\!-23$.
}
\label{fig:difflf}
\end{figure*}

\subsection{Stacked luminosity functions}

In Fig.~\ref{fig:cumulf}, we show the cumulative $B$-band 
luminosity function \index{luminosity
function} (LF) for our optical subsample (25 GEMS groups), evaluated by
co-adding galaxies of several groups in equally spaced bins of
absolute luminosity, with galaxies chosen from within $0.3\,R_{500}$
from the centre of each group.  This reveals that the LF of the X-ray
dim groups ($L_X \!<\! 10^{41.7}$ erg~s$^{-1}$) is significantly
different from that of the X-ray brighter groups.
Fig.~\ref{fig:difflf} shows the differential LFs, (left) for the
$B$-band, within $0.3\,R_{500}$ (same data as Fig.~\ref{fig:cumulf}),
and (right) for the $K$-band, within $R_{500}$. The differential LFs
reveal the nature of the difference, in the form of a 
prominent dip,  at around $M_B \!=\!-18$
and $M_B \!=\!-23.5$ (more details of  observational data
in \cite{miles04,miles06a,miles06b}).

We interpret this deficiency of intermediate luminosity galaxies as
evidence of rapid evolution through merging.  In the low velocity
dispersion environment, as in X-ray dim groups, dynamical friction
\index{dynamical friction} would facilitate more rapid merging, thus
depleting intermediate luminosity galaxies to form more giant central
galaxies.  Since the collision cross-section depends on the size of a
galaxy, the dwarf galaxies at the faint end of the LF are more likely
to merge with a giant galaxy, than merge with each other. This ensures
galaxies in a range of intermediate luminosities are preferentially
depleted, thus enhancing the bright end of the LF.

We suggest that X-ray dim (or low velocity dispersion) groups are the
present sites of rapid dynamical evolution 
\index{dynamical evolution} rather than their X-ray
bright counterparts, and may be the modern precursors of fossil
groups\index{fossil groups}.  
Such LF features are seen in some clusters as well, such as
Coma (e.g. \cite{tully05}), which we would argue has resulted from recent
merger with groups.

\begin{figure*}[tbh!]
\centering
{\includegraphics[height=5.4cm,angle=-90]{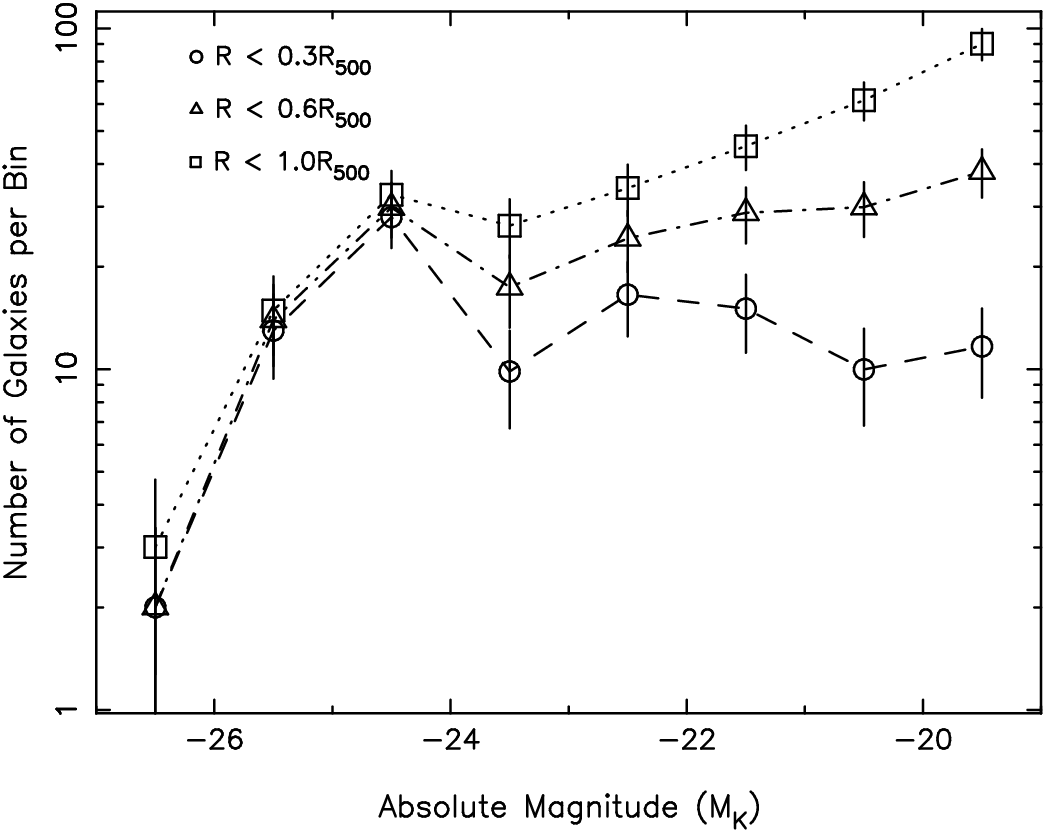}
\includegraphics[height=5.4cm,angle=-90]{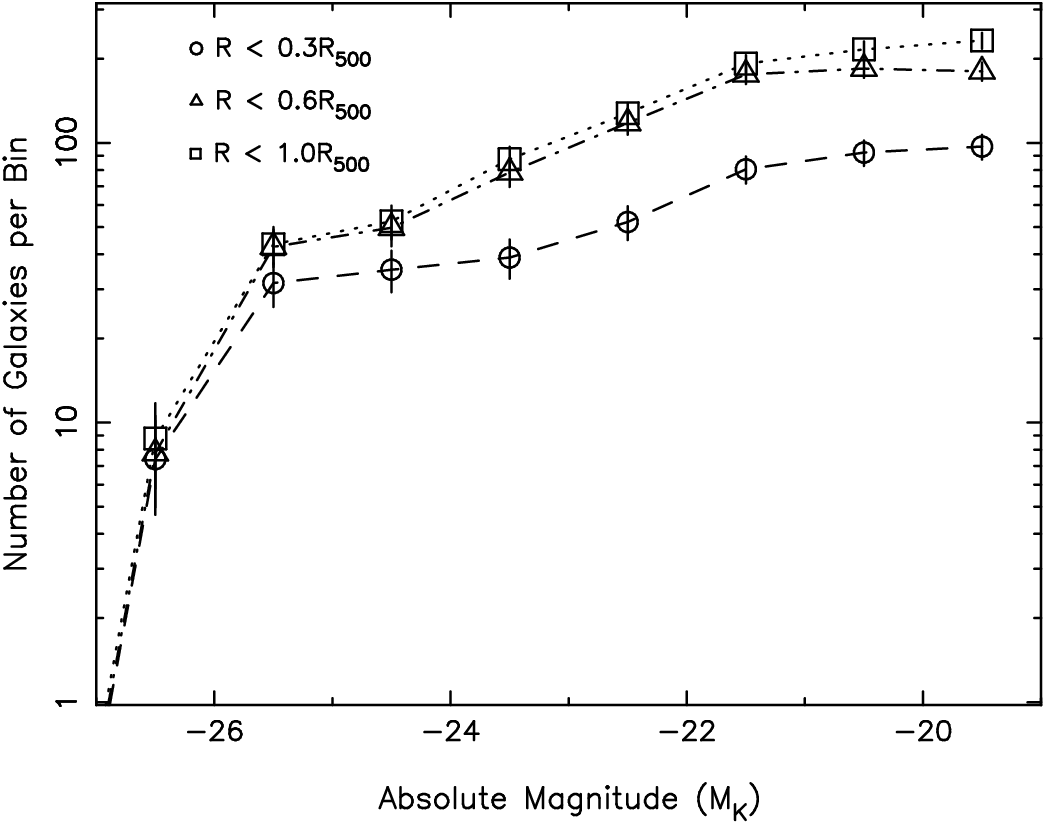}}
\caption{Differential $K$-band luminosity functions of 
{\it (Left:)} all
X-ray dim GEMS groups ($L_X<10^{41.7}$ erg~s$^{-1}$), out to a
fraction of the projected group radius.
{\it (Right:)} The same for all X-ray bright
groups ($L_X>10^{41.7}$ erg~s$^{-1}$). The three plots in
each case go out to 0.3, 0.6 and 1.0 times $R_{500}$ respectively.
The intermediate luminosity dip feature is more prominent in the inner
regions of the X-ray dim groups. The LF of the X-ray bright groups remains
similar in shape at all radii.}
\label{fig:lfrad}       
\end{figure*}

The optical LFs are determined only out to $0.3\,R_{500}$ of each
group, but we can investigate the nature of the LF in their outer parts
from the near-IR LFs from the 2MASS survey.  We consider
three ranges of radial distance in finding the 
the mean $K$-band LF in Fig.~\ref{fig:lfrad}.  The dip between
$-24\!<\! M_K\!<\! -23$ seen in the LF in the central regions of the groups
gradually disappears as the LF is averaged out to larger radii,
approaching $R_{500}$, where the LFs of
both X-ray bright and dim groups are of a similar shape.

\subsection{Brightest group galaxies}\index{brightest group galaxies} 

\begin{figure*}
\centering
{\includegraphics[height=5.4cm,angle=-90]{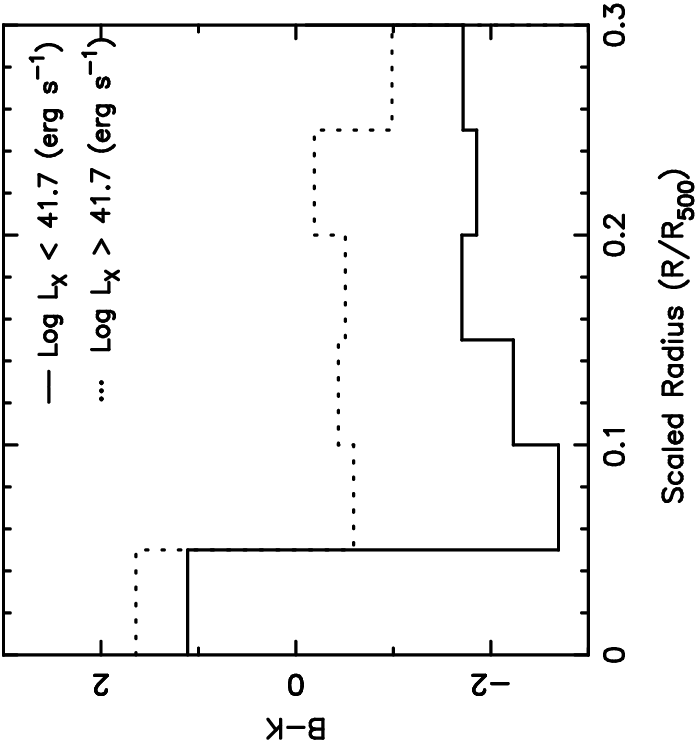}
\includegraphics[height=4.0cm,angle=-90]{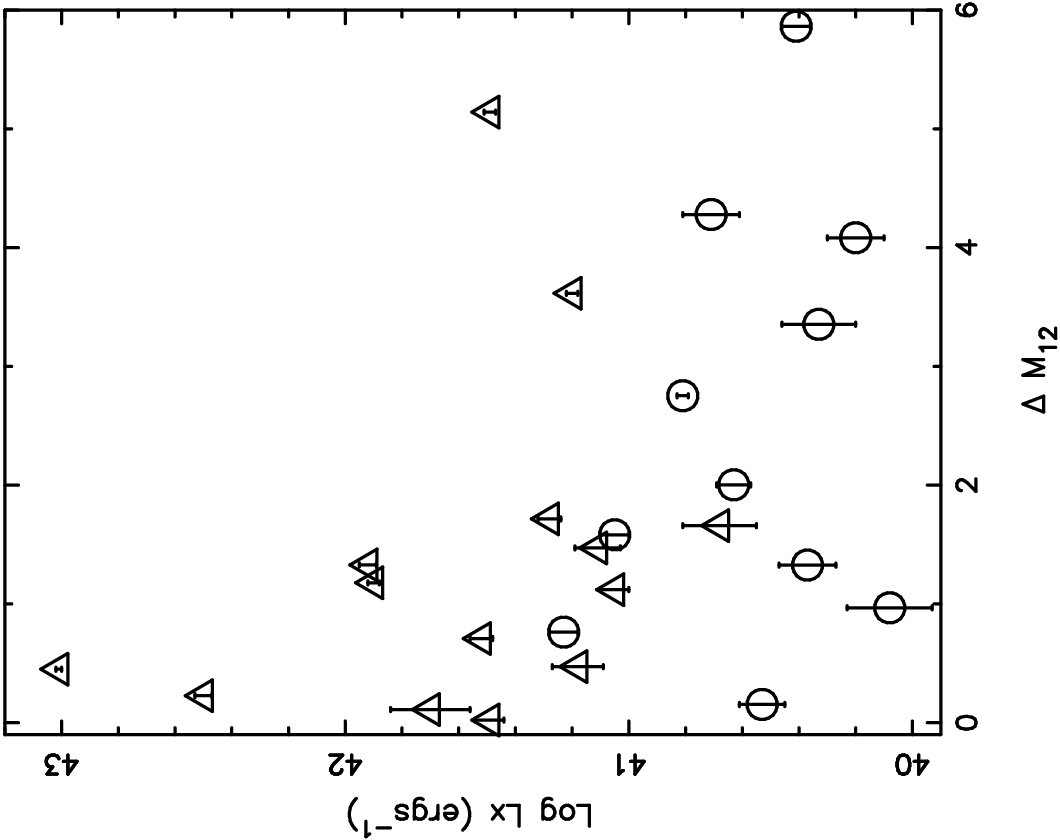}}
\caption{{\it Left:} The average $B\!-\!K$ 
colour of galaxies as function of 
distance (scaled by $R_{500}$) from the centre of the parent group.  
X-ray dim groups have redder central galaxies.
{\it Right:} The X-ray luminosity of our GEMS groups, as a function
of the difference in magnitude between in the brightest and
second brightest galaxies in each group. The biggest values for
$\Delta M_{12}$ are in the X-ray dim groups. 
}
\label{fig:Lxmm2}       
\end{figure*}

One of the consequences of this scenario is that the brightest group
galaxies in the X-ray dim groups are expected to be more massive and
brighter than those in X-ray bright groups.  Fig.~\ref{fig:Lxmm2}
shows the colour of galaxies as a function of radial distance, stacked
in radial bins scales by $R_{500}$, to reveal that X-ray dim groups
have redder central galaxies.  It also shows that in X-ray bright
groups, the difference in $B$-magnitude between their brightest and
second brightest galaxies is in general smaller than in X-ray faint
groups. The X-ray bright groups have several galaxies of comparable
luminosity (and mass) at the bright end, possibly being the end-products of
earlier mergers on smaller scales in sub-groups that were incorporated
in the virialised systems we observe today.

\section{Conclusions}
We argue that the missing intermediate-luminosity galaxies in the
optical and near-infrared luminosity functions of X-ray dim groups
indicate that, in the dynamically sluggish environment of such groups
(which have low velocity disperison), dynamical friction would
facilitate more rapid merging, thus depleting intermediate-luminosity
galaxies to form a few giant central galaxies.  We also show that this
effect is seen only in the interior regions of the groups
($R<0.3\,R_{500}$), and vanishes as one approaches $R_{500}$. 
rather than a bright-end enhancement caused by excess star formation.
In \cite{miles06a}, we show that this feature cannot arise due to
enhanced star formation in the brightest galaxies, or due to a varying
morphological mix of galaxies in various groups.

It is often suggested (e.g. \cite{con06}) that mergers 
\index{mergers} are not an
important ingredient of galaxy evolution in the recent Universe
($z\!<\!1$). Here we have shown that in nearby poor groups, merging is still
an important process.  This picture of galaxy evolution
leads to a definite prediction.  If X-ray dim groups 
are indeed systems undergoing rapid dynamical
evolution, the stellar populations in their galaxies would be
significantly younger than those in X-ray bright groups in case of
dissipative merging.  This can be observationally verified.

\subparagraph{Acknowledgements} 
We thank Trevor Ponman, Duncan Forbes, Paul Goudfrooij and 
other members of the GEMS team for their contributions to this work,
and the organisers for a very stimulating meeting.

%
%

\printindex
\end{document}